# Effects of growth rate, size, and light availability on tree survival across life stages: a demographic analysis accounting for missing values and small sample sizes.

**Aristides Moustakas[1,*] & Matthew R Evans[1]**

1. School of Biological and Chemical Sciences

Queen Mary University of London

Mile End Road, E1 4NS, London, UK

* Corresponding author

Emails: arismoustakas@gmail.com (Aristides Moustakas); matthew.evans@qmul.ac.uk

(Matthew Evans)

**Abstract**

**Background**

Plant survival is a key factor in forest dynamics and survival probabilities often vary across life stages. Studies specifically aimed at assessing tree survival are unusual and so data initially designed for other purposes often need to be used; such data are more likely to contain errors than data collected for this specific purpose.

**Results**

We investigate the survival rates of ten tree species in a dataset designed to monitor growth rates. As some individuals were not included in the census at some time points we use capture-mark-recapture methods both to allow us to account for missing individuals, and to estimate relocation probabilities. Growth rates, size, and light availability were included as covariates in the model predicting survival rates. The study demonstrates that tree mortality is best described as constant between years and size-dependent at early life stages and size independent at later life stages for most species of UK hardwood. We have demonstrated that even with a twenty-year dataset it is possible to discern variability both between individuals and between species.

**Conclusions**

Our work illustrates the potential utility of the method applied here for calculating plant population dynamics parameters in time replicated datasets with small sample sizes and missing individuals without any loss of sample size, and including explanatory covariates.

**Introduction**

Understanding long-term population dynamics of different life stages and the predominant underlying factors that shape them is a central question in biology. A key process in population dynamics is mortality and understanding the factors that influence rates of mortality both between and among life stages is an important endeavour in population biology [1]. Plants are of particular interest as they are reported to defy the effects of ageing [2] yet they die. Trees are typically extremely long-lived and require long-term studies to allow any understanding of the factors that underlie any variations in mortality between individuals and species. Long-term studies bring their own associated problems, in particular they rely on individuals being relocatable on each occasion that a census is conducted, when any means used to identify individuals is likely to get lost or to become indistinguishable.

While it remains elementary that 'plants stand still to be counted and do not have to be trapped, shot, chased, or estimated'[3], monitoring and analysing long-term datasets often involves accounting for data collected under different experimental design protocols, and/or omissions and errors that may occur at some time points. In plants common causes of omissions of individuals and detection gaps where plants are not recorded in one or more surveys but recorded both before and after, include herbivory and re-growth thereafter, dormancy, and recording error [4]. A potential way to surmount this problem is to use a technique widely used in animal ecology, capture-mark-recapture that provides unbiased estimates of survival rates accounting for imperfect detection rates (for applications in plants see e.g. [4, 5]). Alternative solutions such as excluding individuals with measurment gaps would result in either reducing the sample size of the dataset if such individuals were excluded or inflating mortality rates if an individual was classified as dead as soon as it was not recorded [6].

As trees are typically very long-lived few studies have been initiated that are aimed at measuring survival rates directly. If one was to do so then one would design the study so as to minimise the possible sources of error e.g. [7-9]. More typically a researcher interested in tree survival needs to utilise datasets collected for other purposes, and which have probably been collected over much less than one tree lifespan. Such datasets are less likely to have collected in a manner that aims to minimise the errors important in survival analysis. In this study we use a dataset collected to measure growth rates over a twenty-year period and attempt to determine whether we can estimate survival rates of these long-lived organisms.

Estimating survival in the field is also inherently difficult. This difficulty arises because of the difficulty of separating mortality from a failure to relocate an individual. This problem has received a substantial amount of attention in the animal ecology literature, where the probability of recapturing or resighting an individual can be relatively low [10, 11]. It is a tautology that not all individuals that are marked at the beginning of a study will inevitably be relocated when the next census is taken [12]. Even on large, static organisms like trees marks fade, get lost, or simply overlooked. Survival estimation therefore needs methods that account for the relocation probability independently of survival estimation.

Tree survival rates, or reciprocally mortality rates, are important for foresters and conservationists, urban planners, as well as for biologists aiming to understand death as an ecological process [13, 14]. The most common way to quantify and compare mortality rates of tree populations is to measure death counts over a time interval [3]. While mortality rates i.e. the proportion of individuals within the population surviving within the next time unit are interesting *per se*, ecologists have strived to correlate tree death with potential explanatory covariates with the most notable being light availability, growth rates, indices of tree size, climatic variables, as well as density effects [15-19].

Light availability is a key factor in forest dynamics as interspecific differences in shade tolerance underpin species coexistence and species succession in forests [20, 21]. Light is often the limiting factor in tree survival or growth in both northern hemisphere hardwood forests and tropical rainforests - tree individuals exhibit plasticity in terms of tree architecture to utilise light resources and allocate then into growth and survival [22]. Plasticity of canopy architecture for maximising light resources may also be seen as asymmetric competition for light [23] and can explain the evolution of height [24]. Apart from plasticity, competition for light affects tree demography in terms of both mortality and recruitment [25]. The role of light is further emphasized by findings reporting that under low light, small differences in tree growth at early life stages result in variation in rates of mortality of at least two levels of magnitude between species [26]. The effect of light availability across tree life stages is not homogeneous; species with low mortality in the absence of light at early life stages may have slow growth rates at intermediate or late life stages [21]. Therefore, we hypothesize that inclusion of the effect of light will improve predictions of plant survival and that this effect will be more pronounced at early life stages (saplings) than at later life stages. We also suggest that at later life stages (adults) the effect of light will be more pronounced on the survival of canopy species rather than sub-canopy species.

The size of individuals is critical for survival in plants; self thinning trees that are small in comparison to their neighbours are usually the least successful in encapsulating resources and ultimately in surviving [27]. Large trees are more likely to have higher access to light. Size is also important in modulating shade tolerance as there is an inverse relationship between shade tolerance and tree size between species as well as between individuals within a species [28]. Within species there is also a relationship between size and survival (or death) as small individuals at early life stages typically suffer high mortality rates from herbivory, mechanical damage, or light availability [14], but they exhibit higher shade tolerance due to the fact that they have higher photosynthetic to non-photosynthetic biomass ratio [29]. Overall, tree death and reciprocally survival may be size-independent within species, with a constant risk of death regardless of the size of individuals [14, 30], or size-dependent mortality peaking at some particular size classes [14, 18, 31], however these distributions usually refer to, and are derived from, adult trees. During earlier life stages, differences in mortalities between species are more pronounced [21] and the probability of a tree reaching the canopy is often determined by its performance as a sapling [26]. Therefore we hypothesized that the inclusion of size as a covariate will improve predictions of survival and that this will be more pronounced at earlier life stages than at later life stages.

Growth rates of trees are key factors for species composition in forests for several reasons: There is a trade-off between low light survival and high light growth that promotes succession - species that grow faster in full light exhibit lower shade tolerance [32] and lower variation of growth rates [33]. Further there is evidence that tree growth is also determined by asymmetric competition for light [34, 35] and thus the effects of light should be accounted for when calculating growth rates. In addition, increased growth rates may decrease longevities of individuals within species [36]. Growth rates have been used as predictors of tree death [37] and negative growth rates may provide an indicator of forthcoming tree death [38]. We hypothesized that the inclusion of growth rates will improve predictions of survival analysis and this effect will be more pronounced at earlier life stages across all species.

In this study we analyse survival rates of tree species distinguishing between early life stages (saplings) and late life stages (adults) using long-term and time replicated data collected on ten tree species in a lowland forest in the UK. As some individual trees are likely to have been missed at some censuses we have applied a commonly used method in animal ecology, capture-mark-recapture to account for the missing data. Specifically we use a Cormack-Jolly-Seber (CJS) model to estimate both the relocation probability and the survival rate of each tree species. We further included size, growth and light as potential explanatory covariates in the model determining survival rates of species. We have developed a set of models (table 1) that would potentially fit the data regarding survival at a species level. After selecting for the most parsimonious model for each species and life stages, we calculate the probability of survival.

There are potential competing ecological hypotheses that might explain variation in growth rate between species. Firstly, survival rate might scale vary with life history so that trees might 'live fast and die young', we might therefore predict that variation in growth rates between species might be related to variation in survival rate – such that species with high growth rates had low survival rates and vice versa. Secondly, we might expect shade tolerant trees to have relatively high survival when small compared to shade intolerant trees but that the difference in survival rate between shade tolerant and intolerant trees would be negligible when they were large (and in the canopy). Of the tree species in our data set Oak, Beech, Ash and Sycamore are classified as shade tolerant canopy species while Birch and Willow are usually classified as shade intolerant early succession species and Hazel, Field Maple, Elder and Hawthorn as shade tolerant sub-canopy species [39-44]. Therefore, we might expect the Oak, Beech, Ash and Sycamore to have higher survival as small sub-canopy trees than Birch and Willow; for Hazel, Field Maple, Elder and Hawthorn to have survival rates that are vary little with size as they are always sub-canopy.

**Methods**

Study site and species

We used a dataset comprising of 281 individual trees from the Environmental Change Network, Wytham Woods (ECN-W), in Oxfordshire, UK long term monitoring plots. Wytham Woods is located at 51° 46′ N, 1° 20′ W, altitude ranges between 60 - 165 m above sea level, comprised of a surface area of 400 ha, with mean annual temperature 9.9 $^{\circ}$C y$^{-1}$ and mean annual precipitation 728 mm y$^{-1}$ (Sykes & Lane 1996). A network of 10 m x 10 m forest-monitoring plots is distributed across the site (41 plots in total). Tree species monitored included 10 species: *Acer pseudoplatanus* (Sycamore, ACERPS, $N_{saplings}$ = 16, $N_{adults}$ = 41), *Fraxinus excelsior*, (European ash, FRAXEX, $N_{saplings}$ = 16, $N_{adults}$ = 41), *Quercus robur* (Pedunculate oak, QUERRO, $N_{saplings}$ = 5, $N_{adults}$ = 18), *Fagus sylvatica*, (European beech, FAGUSY, $N_{saplings}$ = 5, $N_{adults}$ = 19), *Corylus avellana* (Common hazel, CORYAV, $N_{saplings}$ = 18, $N_{adults}$ = 11), *Crataegus monogyna* (Common hawthorn, CRATMO, $N_{saplings}$ = 8, $N_{adults}$ = 13), *Acer campestre* (Field maple, ACERCA, $N_{saplings}$ = 0, $N_{adults}$ = 13), *Betula* spp. (Birch, BETUSP, $N_{saplings}$ = 0, $N_{adults}$ = 13), *Salix* spp. (Willow, SALISP, $N_{saplings}$ = 4, $N_{adults}$ = 4), and *Sambucus nigra* (Elder, SAMBINI, $N_{saplings}$ = 12, $N_{adults}$ = 11). The encounter raw data used in this study can be found at http://dx.doi.org/10.5061/dryad.6f4qs.

Data description

Initially, ten individuals were marked in each plot. However, as some individuals could not be relocated over time (its often unclear whether they died or whether the mark disappeared), the missing tree was replaced by the nearest unmarked individual within the plot to maintain 10 individuals per plot. Trees were first marked in 1993 and had their Diameter at Breast Height (DBH) measured, attempts were made to relocate and remeasure trees in 1996, 1999, 2002, 2005, 2008 and 2012. Only live trees were measured on each occasion. This time series therefore represents a record of tree survival and mortality as well as their change in size. That relocation of trees was not perfect is exemplified by the fact that some trees were measured in one time period, missed in the next but relocated and remeasured at some point in the future. Clearly, therefore the probability of relocating a tree needs to be taken into account when calculating survival rates otherwise failure to relocate will inflate mortality rates [6]. Therefore, we have analysed the survival time series using capture-mark-recapture analysis that estimates relocation probability as well as survival [45]. Capture-mark-recapture has been applied in plants in order to account for similar problems to the ones faced here [4-6]. In addition we have examined the effect of an individual's size in terms of mean DBH, its mean growth rate (mean annual change in DBH) over the 19 year period and light intensity (expressed as % of full light) measured at each tree in 2012.

DBH was measured on each tree every time it was relocated and when it entered the census. DBH is the tree's diameter measured at a standardised 1.35m above ground level and was estimated by measuring the tree's circumference and diameter estimated using the formula diameter = circumference / π. Growth rate was calculated over each sampling interval (change in DBH / number of years between sampling points), this provided a maximum of six growth rates for each individual tree and the average of all the available growth rates for a tree was used as the mean growth rate for that individual.

Light meter readings were taken on cloudy days during September 2012 at three positions within 1m of the trunk of each tree at a height of 1.35m and simultaneously in a large open gap nearby. To measure light levels we used two PAR Quantum sensors (SKP215,

Skye Instruments Ltd, Llandrindod Wells, UK) which were both calibrated against the same reference lamp, the one used to measure light levels under the canopy was used with a meter (SKP 200, Skye Instruments Ltd, Llandrindod Wells, UK) recording to one decimal place, the one measuring light levels in the open was used with a datalogger (SDL5050 DataHog 2, Skye Instruments Ltd, Llandrindod Wells, UK). Measurements from the sensor in the open gap were made every 10 s with the mean of these more frequent values recorded every 10 min. The data used in the analysis was the proportion of the available light that reached each tree's position, this was estimated by dividing each measurement taken under the canopy by the open gap value which was taken closest in time (always within 10 mins) to the under canopy measurement, this gives three light intensity values for each tree which were averaged to produce a single value for each individual tree [33]. Trees were coded as to whether they were saplings (<10 cm mean DBH) or adult trees (≥10 cm DBH).

Estimation of survival rates.

The data were analysed using a commonly used capture-mark-recapture software package MARK, which applies a modified Cormack-Jolly-Seber model parameterised for relocation probability and survival rate [46]. Capture-mark-recapture methods assume that there are a number of sampling occasions on which individuals are marked in some manner. On each subsequent sampling occasion marked individuals can be relocated and their mark recorded, and unmarked individuals may be marked for the first time. CJS models make several assumptions [11, 46], which need to be tested when a model is run:

1. all marked individuals present in the population at time ($i$) have the same probability of relocation ($p_i$)
2. all marked individuals in the population after time ($i$) have the same probability of surviving to the next sampling occasion at time ($i + 1$)

3. marks are not lost or missed
4. all sampling occasions are instantaneous, relative to the interval between sampling occasions

CJS models use encounter histories of individuals (an encounter history is simply a record of the sampling occasions on which an individual was recorded e.g. 1010 would signify that the individual was marked on the first sampling occasion, it was not relocated on the second sampling occasion but was seen on the third but not the fourth) to estimate both the probability of individuals surviving between sampling occasions and of being relocated on a particular occasion. The basic method can be understood if one follows that the encounter history 111 would have an encounter history probability of ($\varphi_1 p_2 \varphi_2 p_3$, or the probability of surviving the first time interval * probability of relocation at sampling occasion 2 * probability of surviving the second time interval * probability of relocation at sampling occasion 3) while the encounter history 101 would have the probability of ($\varphi_1 (1 - p_2)\varphi_2 p_3$, or the probability of surviving the first time interval * probability of not being relocated at sampling occasion 2 * probability of surviving the second time interval * probability of relocation at sampling occasion 3) and so on.

There are sufficient numbers of trees in our dataset to allow both the relocation probability and the survival rate to be estimated for each species. The models had six unequal time intervals (1993-1996, 1996-1999, 1999-2002, 2002-2005, 2005-2008, 2008-2012) trees were mainly marked in 1993 but some were added at each subsequent time point. In all models the annual relocation probability (p) was kept constant as initial data exploration indicated that there was no *a priori* year-specific bias on the relocation probability. We compared a series of models for each of the ten tree species for which sufficient data existed. This was done twice, firstly distinguishing between saplings and adults and secondly treating all trees as equivalent. A logit link function was used for all models that included covariates, a sine link function for models without covariates [46]. Akaike's information criterion (AIC) was used to select the model that most closely described the data [47].

The suitability of using a CJS model was tested by generating a fully parameterised CJS model (in which both survival and relocation probability varied between years) for each tree species and then performing a parametric bootstrap [48] within MARK [49]. This used the parameter estimates of the model to simulate data that exactly meet the assumptions of a CJS model (no overdispersion, all individuals independent of one another and no violations of model assumptions [11]). We then tested the goodness of fit of the model to the data for each species and estimated the quasi-likelihood parameter ($\hat{c}$) for each species. The species specific $\hat{c}$ was then used to correct for overdispersion in the data. The survival models for each species were re-run using the species specific $\hat{c}$.

Relations between the growth, DBH and light environment of trees and their probability of survival.

We added covariates to the survival models to test for the influence of growth rates, DBH and light environment on the probability of survival. These three factors were added as independent factors and an intercept term was included in all models. Model selection used the corrected AIC (AICc) as described above with the species specific $\hat{c}$ obtained from the bootstrapping exercise used to correct for the degree of overdispersion in the data [49, 50], therefore are correctly referred to as quasi AICc (QAICc).

Survival models

*a) Distinguishing between saplings and adults.*

We defined 17 models (table 1). The age of the tree was included as an attribute of the individual tree (sapling or adult). The effects of growth rate, DBH and light conditions were included for saplings and adults independently and in combination as described in table 1. In most models survival probability was constant between years, in two models we allowed survival probability in the last time period to differ from the earlier time periods.

*b) Treating all trees as equivalent*

In the previous analysis we found that sapling DBH was an important factor in most tree species. As the distinction between sapling and adult is based on DBH, we decided to develop a series of simpler models which treated all trees as equivalent (i.e. we lost the distinction between sapling and adult) but in which we retained DBH as a covariate in all models. Survival probability was kept in constant between years in all models. The set of models used in this category is shown in table 2.

**Results**

<u>Goodness of fit testing</u>

Fully parameterised CJS models ($\varphi_t p_t$, i.e. both survival and relocation vary with time) were used in the bootstrapping exercises, with 1000 simulations performed. The model deviances, the range of bootstrapped deviances and the estimated ^c (observed deviance/expected deviance) for each species are shown in Supplement 1a.

Further analysis of the data using program RELEASE bundled with MARK revealed that test 2 showed significant departures from model assumptions, this means that there was heterogeneity in the probability of relocation. This analysis suggests that the probability of an individual being relocated on occasion ($i + 1$) was more likely if they had been recorded on occasion ($i$) than would be expected by chance ($\chi^2 = 122.45$, d.f. = 4, $P < 0.0001$). In contrast the lack of significance of test 3 suggests no significant departures from the assumption that all individuals alive at occasion ($i$) are equally likely to survive to occasion ($i + 1$).

As the fully parameterised model broke a key assumption of a CJS model (as demonstrated by the significant effect found for test 2), we tested a simplified model in which the survival of individual trees was constrained to be constant in all time intervals but was free to differ between adults and saplings. The relocation probability was also constrained to be constant for all individuals for all occasions. The output of the bootstrapping exercise on the simplified model to determine model goodness of fit and to estimate the quasi-likelihood parameter used in subsequent analyses of survival for each tree species are shown in Supplement 1b.

For seven of the ten tree species the simplified model did not violate the assumptions of a CJS model. In all cases the simplified model met the assumptions of a CJS model better than the fully parameterised version and we decided to proceed with the analysis with a model of this structure. The degree of overdispersion was corrected for by adjusting the value of $\hat{c}$ for each species separately.

Model Selection.

The output of the top three of the 17 models generated for each species are shown in table 3 & supplement 1c. The same three models were the best fitting models for every tree species, with model 2 ($\varphi$ c sapling DBH adult c p c) or model 1 (adult c p c) when no saplings were recorded, being the best fitting model in seven of the ten species.

As the classification of a tree as a sapling or an adult is based on DBH (saplings < 10cm ≥ adults) and the model that most commonly provided the best fit to the data was one that involved DBH, we ran a further set of models in which the sapling – adult classification was dropped and DBH was constrained to remain in the model. The output of these models is shown in table 4. These analyses show that the top two models for all tree species are models one and four, i.e. annual survival is better predicted by DBH alone or by all three covariates in combination.

Parameter estimation.

*Treating saplings and adults separately.*

Models 1 and 2 have the lowest QAICc for all tree species except SALISP, for which they were second and third lowest QAICc. The model estimates derived from these models are shown in Fig. 1 and table 5. These results derive from model 1 in table 1 – constant survival rates with time, constant relocation probability with time, no individual level covariates These analyses suggest that survival varies with DBH in saplings and not in adults. SALISP was the only species for which the model with time varying survival was a better fit than constant survival. The different estimates are:

|  |  | Mean survival probability 1993-2008 | Mean survival probability 2008-2012 | St Error 1993-2008 | St Error 2008-2012 |
|---|---|---|---|---|---|
| Sapling | SALISP | 0.93 | 0.76 | 0.032 | 0.201 |
| Adult | SALISP | 0.96 | 0.005 | 0.038 | 44.61 |

In saplings the highest survival rates were recorded for Ash and Sycamore, while the lowest were recorded for Willow, Elder, and Hazel (Fig. 2). The highest survival rates for adults were recorded for Beech, Ash, and Sycamore (Fig. 3). The lowest survival rates were derived for Willow, Elder, and Birch (Fig. 3). The overall effects demonstrate strong effects of DBH on sapling Beech and Hazel and rather less noticeable effects on Elder and Willow, with much weaker effects on the other six species (Fig. 2 & Fig 3).

*Treating all trees as equivalent*

Model 1, which includes a DBH effect on all trees was the best fitting model for all tree species except for Beech, in which it was the second best fitting model. Models 1 and 4 were the best fitting models for all tree species. Estimates of survival as a function of DBH from model 1 – constant survival rates with time, constant relocation probability with time, an effect of DBH on survival are listed in table 6. As a logit link function was used in the analysis of covariates the general function relating the probability of survival ($P_{survival}$) to DBH is given by:

$$P_{survival} = \left\{ e^{\left(\beta 1 + \left((\beta 2\ (DBH - \overline{DBH}))\ s_{DBH}\right)\right)} \right\} \div \left\{ 1 + e^{\left(\beta 1 + \left((\beta 2\ (DBH - \overline{DBH}))\ s_{DBH}\right)\right)} \right\}$$

The values of $\beta_1$ and $\beta_2$ are given for all tree species in table 6, $\overline{DBH}$ is the mean DBH of the species concerned and $s_{DBH}$ is the standard deviation of DBH for the species concerned. Model 4, the second best model, included all three covariates (growth rate, light, and DBH) on survival and model outputs are provided in supplement 2a.

This analysis suggests that the survival probability of trees changes with DBH as shown in Fig. 4 (for a detail of Fig. 4 plotted on a more detailed scale see supplement 2b). This demonstrates that there are strong positive effects of DBH on survival in small

Sycamore, Oak and Willow and weaker effects in Birch and Ash. There also are discernable negative size effects on large Beech, Field Maple, Hazel, Hawthorn and Elder.

The rank order of growth rates of adult trees rank order is Hazel, Sycamore, Beech, Oak, Field Maple, Birch, Hawthorn, Ash. The ranking of growth rates of 5cm DBH trees in 10% light is Hazel, Hawthorn, Beech, Ash, Oak, Sycamore; at very low light (1%) order changes to Ash, Beech, Hazel, Hawthorn, Sycamore, Oak (ref to scientific data paper). These figures give a spearman rank correlation of $r = 0.15$ (df = 6, P = 0.72) for adult growth rate versus adult survival and $r = 0.59$ (df = 4, P = 0.21) for saplings in low light and $r = 0.98$ (df = 4, P < 0.001) for saplings in mid light conditions.

**Discussion**

Using a twenty-year dataset that was collected to measure tree growth rates we successfully estimated annual survival probabilities for the ten tree species under consideration. Although sample sizes of either individuals or years were not large (numbers of individuals ranged from 8 to 57) survival rates were estimated with relatively low estimated error. This suggests that the estimation of tree survival rates for population models can be achieved using what might be regarded as suboptimal data. The fact that we were additionally able to estimate the effects of covariates suggests that useful information can be produced from such data.

In our analysis, when the life stages are separated into saplings and adults then DBH appears to affect survival of saplings for five of the eight species (no saplings were recorded for Maple and Birch). Survival of Sycamore and Elder, were both best explained by model 1, a constant survival probability between years without any significant effects of size, growth or light on either life stage. Survival of Willow was best explained by model 16, implying that survival probability varied two periods – pre-2008 and post 2008 but not between life stages and without any of size, growth, and light having a significant effect. If the distinction between saplings and adults is relaxed - as saplings and adults are distinguished by their size in terms of DBH - the best fitting models contained an effect of DBH on survival for nine out of 10 species. For the remaining species, Beech, survival was better predicted by all three covariates, size, light, and growth, in combination. Therefore, even for Beech size had an effect of survival. Our analysis suggests that survival varies with DBH in the species considered here. Examination of Fig. 4 shows that when the coefficients are considered then there are strong positive effects of DBH on survival in small Sycamore, Oak and Willow and weaker effects in Birch and Ash. There also are discernable, but small, negative size effects of size on large Beech, Field Maple, Hazel, Hawthorn and Elder. The two sets of analyses are broadly consistent with each other and suggest that the strongest effects of size are on small/young trees and in general larger trees survive better than small ones. Generally speaking there is little effect of size on large/adult trees although it is possible to demonstrate a small negative effect in some species (Fig. 4).

Effects of growth rates, light, and size on survival

Our analyses suggest that survival is not significantly dependent upon light or growth in Wytham Woods this is not to deny that these factors play an important role in plant physiology and ultimately population dynamics (see also supplement 2a). Trees are more likely to become large when growing quickly and/or when they have access to high light resources [51, 52]. Our results imply that a size and not light or growth related selection process is likely to be acting at early life stages and the intensity of this process is acting at a similar magnitude between years (i.e. survival probability was best fitted as constant between years). Once becoming adult, survival shows there is a smaller effect of size and little evidence of any effect of light, or growth. This is partly in agreement with recent studies in tropical and Mediterranean forests reporting that tree size was the most important predictor of tree survival, followed by biotic and then abiotic variables [16]. It should be noted however that while size and growth covariates were available during all intervals of the study, light availability was only available during the last time interval. Such analysis has been conducted in other studies too [33] but we cannot infer the potential difference in the availability of light in previous time steps. In our study the effect of size was most noticeable at early life stages indicating that in our study site interspecific differences in sapling mortality are going to be key components of forest community dynamics as reported in US hardwoods [26, 53]. Our results are in agreement with metabolic theory predicting that tree mortality rates should decrease with tree size, and tree mortality should scale with tree diameter with a constant exponent [54].

Survival rates and species demographics

Unsurprisingly our results show that trees have high survival rates, but we have been able to demonstrate even with a twenty year dataset that it is possible to discern some variability both between individuals (as already discussed) and between species. Willow appears to behave differently from the other species considered here – it is the only species for which a model that had a non-constant survival probability was a better fit than a constant survival, this suggested that the annual survival probability post-2008 was much lower than that from 1993-2008. Willows are relatively rare at Wytham and it may be that this wood is simply poorly suited to their requirements. It is also worth noting that for many of the species examined the relocation probability was less than one. For some species it was considerably less than one, which suggests that the chance of an observer failing to relocate an individual tree when it was present could be almost as high as one in ten for some species.

In terms of our ecological hypotheses in general the sub-canopy species (Maple, Birch, Hazel, Hawthorn, Willow and Elder) have slightly lower mean survival rates than the canopy species (Sycamore, Ash, Oak and Beech) supporting the hypothesis that these canopy species are correctly classified as shade tolerant. In three out of four canopy species (Ash, Oak, Beech) survival rates were higher for adults than sapling. However, Sycamore exhibited slightly higher survival rates for saplings than adults. From the sub-canopy species Hawthorn had equal survival rates for adults and saplings, while Elder had slightly higher survival rates of saplings while Hazel had higher survival rates for adults than saplings.

In terms of our ecological hypotheses we have no significant evidence that trees live fast and die young. If anything when sapling growth rate was positively related to survival and so in that case they live fast and die late. The four shade-tolerant canopy species had relatively high survival rates when small (rank survival at 5cm were Ash 5, Sycamore 9, Oak 8 and Beech 4) compared to the shade-intolerant sub-canopy species (Willow rank 10, Birch 7). The four shade-tolerant sub-canopy trees had the highest survival when small (Field Maple 1, Hawthorn 3, Hazel 2, Elder 6).

On the use of capture-recapture in plants

There are several scientifically rigorous ways to account for tree survival analysis in relation with explanatory variables such as mixed effects models [16] or Bayesian methods [7, 55] but these do not account for imperfect relocation of individuals. Recent findings report that imperfect detection is the rule rather than the exception in plant distribution studies [56] with negative implications for conservation and distribution modelling [57].

The method of applying capture-mark-recapture to plants *per se* is unusual but not novel, often used with plants with cryptic life stages [4, 5, 58, 59]. Here we argue that capture-recapture with covariates provides a robust alterative to survival analysis of plant individuals in time replicated datasets that might contain individuals that could have been missed at some time step also due to sampling error, missing labels etc. Environmental data and time replicated datasets often contain missing individuals either due to herbivory (e.g. deer herbivory [60]) size-related detection difficulties such as when surveying seedlings [61] or dormant plants [4] and accessing study plots due to adverse weather conditions [62]. In general, studies using linear models with data deriving from the same sites are not fully independent and often include a random effect for site [63]. These analyses cannot be run with missing values, and so a subset of the data with no missing values is often created for analysis, which can result in a significant reduction in sample size [64]. The method as applied here may serve as more desirable alternative for calculating survival and also testing the effects of potential explanatory covariates on survival when it includes missing individuals or individuals with some missing observations without any loss on sample size. While we chose to use the capture-recapture method there are also other methods that can estimate survival rates with missing values such as the Kaplan-Meier [65], the proportional hazards model [66], or Bayesian approximation [67].

Sampling design and data overdispersion

Our analysis showed that the data were overdispersed implying a greater variation than expected from the model [55, 68]. Overdispersion in CJS models is often produced by data non-independence or when probabilities of detection and survival vary between individuals [10]. Examples of data non-independence include whether the plot/individual was located and sampled in the previous time step or if there is a spatially explicit acting pattern [69] such as herbivory or human made logging. To our knowledge there is no selective logging activity in Wytham, and while there was a large population of deer, their numbers have been drastically reduced during the past 30 years. Our results are most consistent with observer bias [70, 71] such that individual trees which were known to be present on the previous occasion were more likely to be observed on the current occasion than trees which were present but not observed on the previous occasion. That is to say that individuals that were inadvertently missed in one survey were often more likely to be missed again in the following survey than would be expected by chance. This could easily occur because marks are often renewed on surveys and a missed individual would not have its mark renewed, or if the observer knew which trees were present in the previous survey and only searched for the ones that were seen on the previous occasion. Our analysis suggests that foresters or those conducting surveys should search for trees even if they were missed at a previous time-step rather than assuming they are dead as they contain information that can be used in long-term demographic studies. Failing to attempt to relocate such trees will result in bias and is likely to inflate estimates of mortality rate [72].

Potential additional applications

We show that relatively small, relatively short duration (in comparison to the species life span) datasets can be usefully analysed to show differences in survival rates between species of tree and between individuals. This suggests that data that have been collected for purposes other than survival estimation can be used for this purpose and that it may be easier than might be imagined to discern the effects of parameters of interest on survival. The survival rates per species reported here are of potential use both for forestry and plant conservation as well as for inclusion in predictive models [73-77] calibrated for Britain or Europe (e.g. SORTIE - [20]). Further implications of our work include the potential use of the method applied here for calculating growth as well as population size including explanatory covariates and thus deriving full population dynamics parameters; in the present paper the method was applied to survival but it can be applied to estimate growth and population size (for examples where this has been done see [5, 58, 59].

**Conclusions**

Our work illustrates the potential utility of capture mark recapture method for assessing survival (or death) rates of trees. Survival or mortality rates have been also attributed to growth rates, light availability, and DBH as potential explanatory covariates. We have demonstrated that even with a twenty year dataset it is possible to discern variability both between individuals and between species. The study showcases that tree mortality is best described as constant between years and size-dependent at early life stages and size independent at later life stages for most species of UK hardwood.

**Competing interests**

The authors declare that they have no competing interests

**Authors' contributions**

AM led the manuscript writing. MRE performed statistical analyses and contributed into manuscript writing.

**Acknowledgements**

We thank David Ward for comments on an earlier manuscript draft, and the numerous scientists associated with data collection in Wytham Woods. Comments from three anonymous reviewers improved an earlier manuscript draft.

**Availability of supporting data**

The data set supporting the results of this article is available in the Dryad repository, http://dx.doi.org/10.5061/dryad.6f4qs.

Moustakas, A. and Evans, M. R. (2015) Effects of growth rate, size, and light availability on tree survival across life stages: a demographic analysis accounting for missing values.

**Table 1.** Descriptions of each model used in the survival analysis that distinguishes between saplings and adults.

| Model | | | Description | |
|---|---|---|---|---|
| | | | Survival probability | Relocation probability |
| 1 | φ: c | P: c | Constant between years | Constant between years |
| 2 | φ: c, sapling DBH; adult c | P: c | Constant between years, varies with DBH for saplings only | Constant between years |
| 3 | φ: c, sapling light; adult c | P: c | Constant between years, varies with light for saplings only | Constant between years |
| 4 | φ: c, sapling growth; adult c | P: c | Constant between years, varies with growth rate for saplings only | Constant between years |
| 5 | φ: c, sapling c; adult DBH | P: c | Constant between years, varies with DBH for adults only | Constant between years |
| 6 | φ: c, sapling DBH; adult DBH | P: c | Constant between years, varies with DBH for saplings and adults | Constant between years |
| 7 | φ: c, sapling growth; adult growth | P: c | Constant between years, varies with growth rate for saplings and adults | Constant between years |
| 8 | φ: c, sapling light; adult light | P: c | Constant between years, varies with light for saplings and adults | Constant between years |
| 9 | φ: c, sapling DBH; adult light | P: c | Constant between years, varies with DBH for saplings and light for adults | Constant between years |
| 10 | φ: c, sapling DBH; adult growth | P: c | Constant between years, varies with DBH for saplings and growth rate for adults | Constant between years |
| 11 | φ: c, sapling light; adult DBH | P: c | Constant between years, varies with light intensity for saplings and DBH for adults | Constant between years |
| 12 | φ: c, sapling light; adult growth | P: c | Constant between years, varies with light intensity for saplings and growth rate for adults | Constant between years |
| 13 | φ: c, sapling growth; adult | P: c | Constant between years, varies with growth rate | Constant between years |
|    | DBH                                                  |      | for saplings and DBH for adults                                                                      |                        |
| 14 | φ: c, sapling growth; adult light                    | P: c | Constant between years, varies with growth rate for saplings and light for adults                    | Constant between years |
| 15 | φ: c, sapling growth DBH light; adult growth DBH light | P: c | Constant between years, varies with growth rate, DBH and light intensity for saplings and adults | Constant between years |
| 16 | φ: 93-08 08-12,                                      | P: c | Varies between 1993-2008 and 2008-2012                                                               | Constant between years |
| 17 | φ: 93-08 08-12, sapling DBH; adult c                 | P: c | Varies between 1993-2008 and 2008-2012, varies with DBH for saplings only                            | Constant between years |

**Table 2.** Descriptions of models used in the survival analysis that treated saplings and adults as equivalent but which retained DBH within all models.

| Model | | | Description | |
|---|---|---|---|---|
| | | | Survival probability | Relocation probability |
| 1 | φ: c DBH; | P: c | Constant between years, varies with DBH | Constant between years |
| 2 | φ: c DBH + light | P: c | Constant between years, varies with DBH and light environment | Constant between years |
| 3 | φ: c DBH + growth | P: c | Constant between years, varies with DBH and growth rate | Constant between years |
| 4 | φ: c, DBH + light + growth | P: c | Constant between years, varies with DBH, light environment and growth | Constant between years |

**Table 3,** the top three models by QAICc for each tree species from the set of models (listed in table 1) that treat adults and sapling separately. Detailed statistics can be found in Supplement 1c. No saplings of ACERCA and BETUSP were recorded, and thus for those two species no distinction between adults and saplings could be made.

| Species | Model No. | Model Description |
|---|---|---|
| ACERCA | | |
| | 1 | φ c p c |
| | 16 | φ 93-08, 08-12 p c |
| ACERPS | 1 | φ c p c |
| | 2 | φ c sapling DBH; adult c p c |
| | 16 | φ 93-08, 08-12 p c |
| BETUSP | | |
| | 1 | φ c p c |
| | 16 | φ 93-08, 08-12 p c |
| CORYAV | 2 | φ c sapling DBH; adult c p c |
| | 1 | φ c p c |
| | 16 | φ 93-08, 08-12 p c |
| CRATMO | 2 | φ c sapling DBH; adult c p c |
| | 1 | φ c p c |
| | 16 | φ 93-08, 08-12 p c |
| FAGUSY | 2 | φ c sapling DBH; adult c p c |
| | 1 | φ c p c |
| | 16 | φ 93-08, 08-12 p c |
| FRAXEX | 2 | φ c sapling DBH; adult c p c |
| | 1 | φ c p c |
| | 16 | φ 93-08, 08-12 p c |
| QUERRO | 2 | φ c sapling DBH; adult c p c |
| | 1 | φ c p c |
| | 16 | φ 93-08, 08-12 p c |
| SALISP | 16 | φ 93-08, 08-12 p c |
| | 1 | φ c p c |
| | 2 | φ c sapling DBH; adult c p c |
| SAMBINI | 1 | φ c p c |
| | 2 | φ c sapling DBH; adult c p c |
| | 16 | φ 93-08, 08-12 p c |

**Table 4.** Output from the four models which treat saplings and adults as equivalent. Models ranked by QAICc.

ACERCA

|   | Model | QAICc | Delta QAICc | QAICc weight | Model Likelihood | No. Parameters | QDeviance |
|---|---|---|---|---|---|---|---|
| 1 | φ c DBH p c | 7.29 | 0.00 | 0.46 | 0.94 | 2 | 3.14 |
| 4 | φ DBH growth light p c | 11.61 | 4.32 | 0.05 | 0.11 | 4 | 3.09 |
| 2 | φ c DBH light p c | 39.77 | 32.48 | 0.00 | 0.00 | 3 | 33.46 |
| 3 | φ c DBH growth p c | 39.77 | 32.48 | 0.00 | 0.00 | 3 | 33.46 |

ACERPS

|   | Model | QAICc | Delta QAICc | QAICc weight | Model Likelihood | No. Parameters | QDeviance |
|---|---|---|---|---|---|---|---|
| 1 | φ c DBH p c | 32.13 | 0.00 | 0.36 | 0.81 | 3 | 25.83 |
| 4 | φ DBH growth light p c | 33.38 | 1.25 | 0.19 | 0.44 | 4 | 24.87 |
| 2 | φ c DBH light p c | 184.78 | 152.65 | 0.00 | 0.00 | 3 | 178.48 |
| 3 | φ c DBH growth p c | 184.78 | 152.65 | 0.00 | 0.00 | 3 | 178.48 |

BETUSP

|   | Model | QAICc | Delta QAICc | QAICc weight | Model Likelihood | No. Parameters | QDeviance |
|---|---|---|---|---|---|---|---|
| 1 | φ c DBH p c | 31.18 | 0.00 | 0.66 | 1.00 | 2 | 27.03 |
| 4 | φ DBH growth light p c | 33.34 | 2.15 | 0.22 | 0.34 | 3 | 27.03 |
| 3 | φ c DBH growth p c | 75.55 | 44.37 | 0.00 | 0.00 | 3 | 69.25 |
| 2 | φ c DBH light p c | 75.55 | 44.37 | 0.00 | 0.00 | 3 | 69.25 |

CORYAV

|   | Model | QAICc | Delta QAICc | QAICc weight | Model Likelihood | No. Parameters | QDeviance |
|---|---|---|---|---|---|---|---|

| | Model | QAICc | Delta QAICc | QAICc weight | Model Likelihood | No. Parameters | QDeviance |
|---|---|---|---|---|---|---|---|
| 1 | φ c DBH p c | 48.62 | 0.00 | 0.38 | 0.99 | 3 | 42.31 |
| 4 | φ DBH growth light p c | 49.56 | 0.94 | 0.24 | 0.62 | 4 | 41.05 |
| 3 | φ c DBH growth p c | 86.80 | 38.18 | 0.00 | 0.00 | 3 | 80.49 |
| 2 | φ c DBH light p c | 86.80 | 38.18 | 0.00 | 0.00 | 3 | 80.49 |

CRATMO

| | Model | QAICc | Delta QAICc | QAICc weight | Model Likelihood | No. Parameters | QDeviance |
|---|---|---|---|---|---|---|---|
| 1 | φ c DBH p c | 15.56 | 0.00 | 0.75 | 1.00 | 2 | 11.41 |
| 4 | φ DBH growth light p c | 19.88 | 4.32 | 0.09 | 0.12 | 4 | 11.37 |
| 3 | φ c DBH growth p c | 76.20 | 60.64 | 0.00 | 0.00 | 3 | 69.90 |
| 2 | φ c DBH light p c | 76.20 | 60.64 | 0.00 | 0.00 | 3 | 69.90 |

FAGUSY

| | Model | QAICc | Delta QAICc | QAICc weight | Model Likelihood | No. Parameters | QDeviance |
|---|---|---|---|---|---|---|---|
| 4 | φ DBH growth light p c | 27.51 | 0.00 | 1.00 | 1.00 | 4 | 19.00 |
| 1 | φ c DBH p c | 43.88 | 16.37 | 0.00 | 0.00 | 3 | 37.58 |
| 2 | φ c DBH light p c | 143.26 | 115.74 | 0.00 | 0.00 | 2 | 139.11 |
| 3 | φ c DBH growth p c | 155.49 | 127.98 | 0.00 | 0.00 | 3 | 149.19 |

FRAXEX

| | Model | QAICc | Delta QAICc | QAICc weight | Model Likelihood | No. Parameters | QDeviance |
|---|---|---|---|---|---|---|---|
| 1 | φ c DBH p c | 27.77 | 0.00 | 0.70 | 1.00 | 2 | 23.62 |
| 4 | φ DBH growth light p c | 31.64 | 3.87 | 0.10 | 0.14 | 4 | 23.13 |
| 2 | φ c DBH light p c | 258.64 | 230.87 | 0.00 | 0.00 | 3 | 252.34 |

|   | Model | QAICc | Delta QAICc | QAICc weight | Model Likelihood | No. Parameters | QDeviance |
|---|---|---|---|---|---|---|---|
|   | c |   |   |   |   |   |   |
| 3 | φ c DBH growth p c | 258.64 | 230.87 | 0.00 | 0.00 | 3 | 252.34 |

QUERRO

|   | Model | QAICc | Delta QAICc | QAICc weight | Model Likelihood | No. Parameters | QDeviance |
|---|---|---|---|---|---|---|---|
| 1 | φ c DBH p c | 21.61 | 0.00 | 0.65 | 1.00 | 2 | 17.46 |
| 4 | φ DBH growth light p c | 23.65 | 2.04 | 0.23 | 0.36 | 4 | 15.14 |
| 2 | φ c DBH light p c | 91.33 | 69.72 | 0.00 | 0.00 | 2 | 87.18 |
| 3 | φ c DBH growth p c | 94.52 | 72.91 | 0.00 | 0.00 | 3 | 88.22 |

SALISP

|   | Model | QAICc | Delta QAICc | QAICc weight | Model Likelihood | No. Parameters | QDeviance |
|---|---|---|---|---|---|---|---|
| 1 | φ c DBH p c | 39.81 | 0.00 | 0.15 | 0.20 | 3 | 33.51 |
| 4 | φ DBH growth light p c | 41.55 | 1.74 | 0.06 | 0.08 | 4 | 33.04 |
| 3 | φ c DBH growth p c | 47.21 | 7.40 | 0.00 | 0.00 | 2 | 43.06 |
| 2 | φ c DBH light p c | 48.68 | 8.86 | 0.00 | 0.00 | 3 | 42.37 |

SAMBINI

|   | Model | QAICc | Delta QAICc | QAICc weight | Model Likelihood | No. Parameters | QDeviance |
|---|---|---|---|---|---|---|---|
| 1 | φ c DBH p c | 67.70 | 0.00 | 0.82 | 1.00 | 3 | 61.40 |
| 4 | φ DBH growth light p c | 72.17 | 4.46 | 0.09 | 0.11 | 5 | 61.39 |
| 2 | φ c DBH light p c | 92.83 | 25.12 | 0.00 | 0.00 | 3 | 86.52 |
| 3 | φ c DBH growth p c | 92.99 | 25.29 | 0.00 | 0.00 | 3 | 86.69 |

**Table 5.** Estimates of adult survival and sapling survival as a function of DBH from model 2 in table 1 – constant survival rates with time, constant relocation probability with time, an effect of DBH on sapling survival but not adult survival. (As a logit link function was used in the analysis of covariates the general function relating the probability of survival to DBH is given by:

$$P_{survival} = \left\{ e^{\left(\beta_1 + \left((\beta_2 (DBH - \overline{DBH})) s_{DBH}\right)\right)} \right\} \div \left\{ 1 + e^{\left(\beta_1 + \left((\beta_2 (DBH - \overline{DBH})) s_{DBH}\right)\right)} \right\}$$

| Age | Species | $\beta_1$ | St error | $\beta_2$ | St error |
| --- | --- | --- | --- | --- | --- |
| Sapling | ACERCA | N/A | N/A | N/A | N/A |
| Adult | ACERCA | 5.48 | 1.002 | | |
| Sapling | ACERPS | 6.25 | 1.008 | 1.20 | 1.68 |
| Adult | ACERPS | 5.41 | 0.581 | | |
| Sapling | BETUSP | N/A | N/A | N/A | N/A |
| Adult | BETUSP | 3.42 | 0.415 | | |
| Sapling | CORYAV | 7.51 | 2.449 | 2.73 | 1.779 |
| Adult | CORYAV | 22.16 | 3704.18 | | |
| Sapling | CRATMO | 4.98 | 0.706 | 0.43 | 0.765 |
| Adult | CRATMO | 4.70 | 0.710 | | |
| Sapling | FAGUSY | 26.51 | 13.24 | 32.46 | 17.46 |
| Adult | FAGUSY | 42.44 | 0.000 | | |
| Sapling | FRAXEX | 5.56 | 0.523 | -0.40 | 0.346 |
| Adult | FRAXEX | 5.90 | 0.708 | | |
| Sapling | QUERRO | 4.68 | 0.613 | 0.673 | 0.856 |
| Adult | QUERRO | 5.10 | 0.709 | | |
| Sapling | SALISP | 2.46 | 0.378 | 0.07 | 0.426 |
| Adult | SALISP | 2.58 | 0.583 | | |
| Sapling | SAMBINI | 3.41 | 0.347 | 0.11 | 0.447 |
| Adult | SAMBINI | 3.35 | 0.440 | | |

**Table 6.** Estimates of survival as a function of DBH from model 1 in table 2 – constant survival rates with time, constant relocation probability with time, an effect of DBH on survival. (As a logit link function was used in the analysis of covariates the general function relating the probability of survival to DBH is given by:

$$P_{survival} = \left\{ e^{\left(\beta_1 + \left((\beta_2 (DBH - \overline{DBH})) s_{DBH}\right)\right)} \right\} \div \left\{ 1 + e^{\left(\beta_1 + \left((\beta_2 (DBH - \overline{DBH})) s_{DBH}\right)\right)} \right\}$$

| Species | $\beta_1$ | St error | $\beta_2$ | St error | Relocation probability | St error |
|---|---|---|---|---|---|---|
| ACERCA | 5.57 | 2.431 | -0.207 | 22.215 | 1.00 | 0.000 |
| ACERPS | 5.62 | 0.967 | 5.23 | 3.456 | 0.99 | 0.009 |
| BETUSP | 3.40 | 0.730 | 0.143 | 2.962 | 1.00 | 0.000 |
| CORYAV | 5.65 | 2.44 | -0.19 | 9.750 | 0.89 | 0.042 |
| CRATMO | 4.83 | 1.49 | -0.09 | 4.763 | 1.00 | 0.000 |
| FAGUSY | 4.68 | 0.998 | -0.227 | 1.759 | 0.99 | 0.012 |
| FRAXEX | 5.36 | 0.929 | 0.44 | 2.115 | 1.00 | 0.000 |
| QUERRO | 4.23 | 1.056 | 1.83 | 5.49 | 1.00 | 0.000 |
| SALISP | 2.67 | 0.678 | 3.30 | 7.289 | 0.90 | 0.084 |
| SAMBINI | 3.44 | 0.841 | -0.08 | 5.315 | 0.89 | 0.015 |

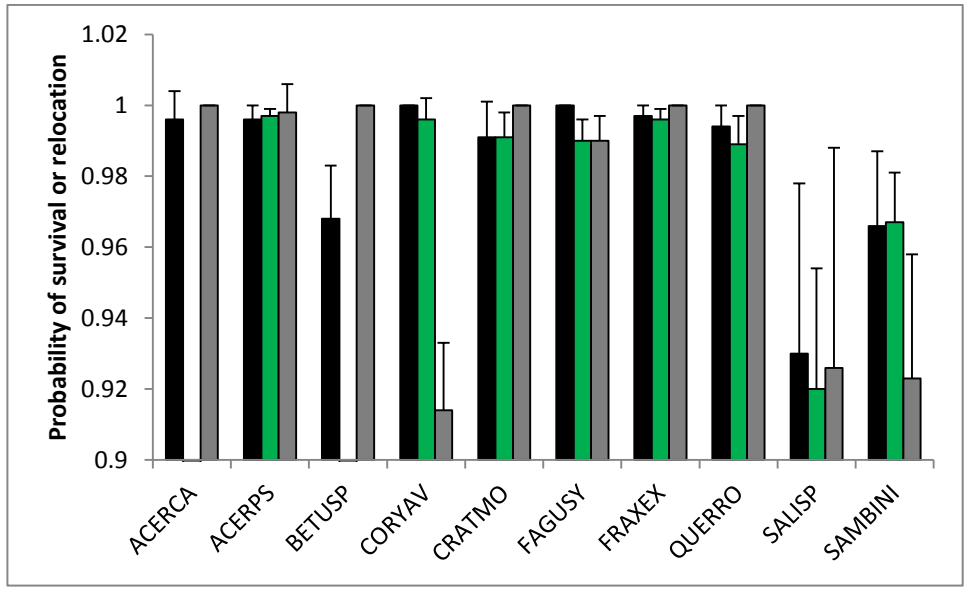

**Figure 1. Annual survival rates and relocation probabilities**

Estimates of annual adult and sapling survival rates and relocation probabilities from model 1 in table 1 – these assume constant survival rates with time, constant relocation probability with time, and has no individual level covariates. Black bars give adult annual survival rates, green bars sapling annual survival rates and grey bars the relocation probability. Error bars give the standard error. SALISP was the only species for which the model with time varying survival was a better fit than constant survival and the optimal model results are listed in the main text.

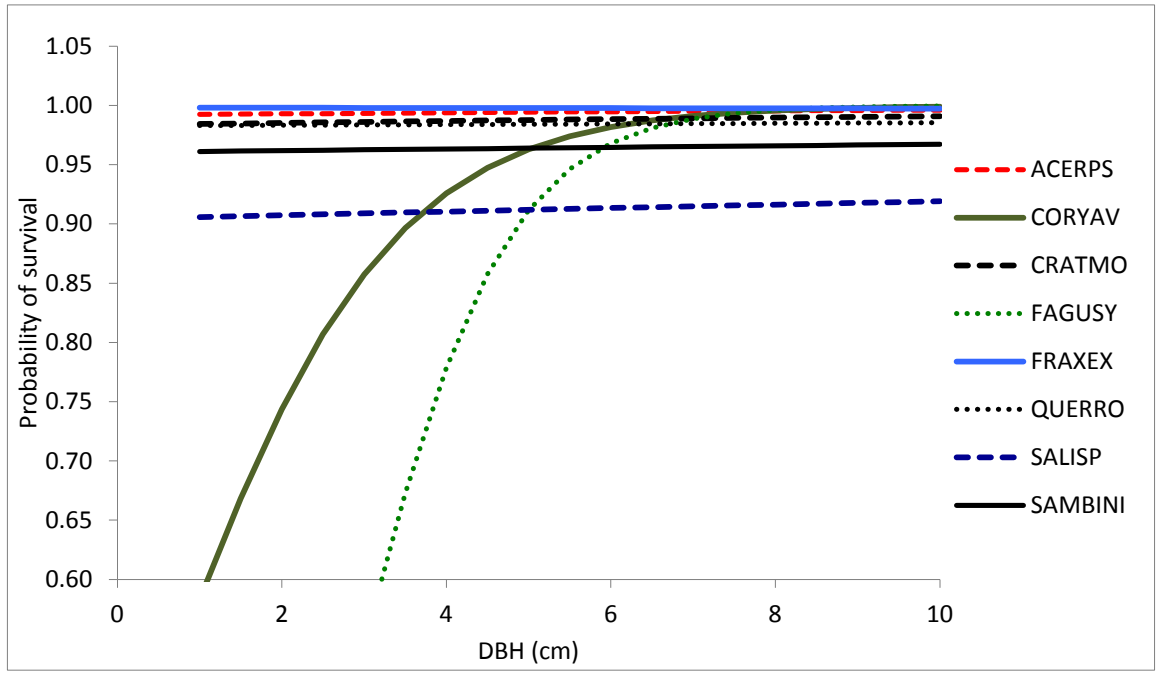

**Figure 2. Survival rates in saplings**

Probability of survival varies with DBH in saplings, derived from model 2 in table 1, which includes a DBH effect on saplings.

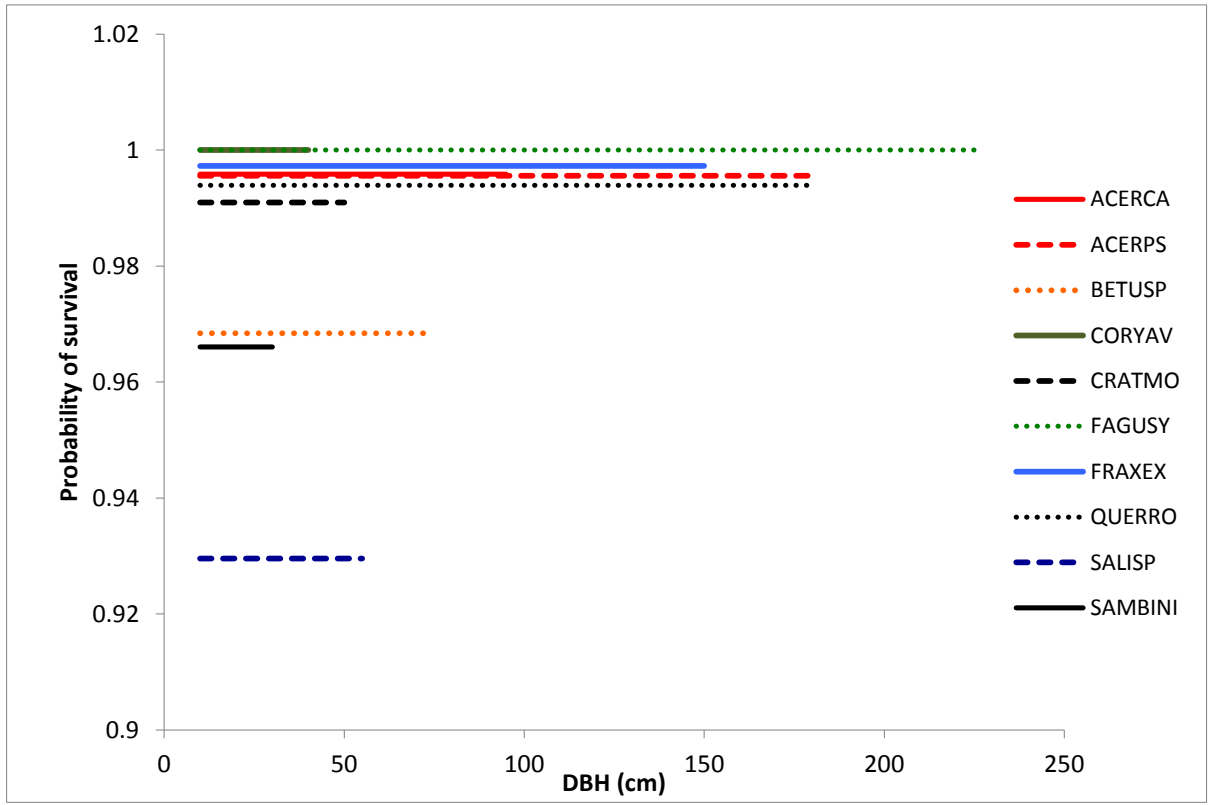

**Figure 3. Survival rates in adults**

Probability of survival does not vary significantly with DBH in adult trees, derived from model 2 in table 1, which does not include a DBH effect on adults. Data are plotted up to the maximum DBH recorded in Wytham.

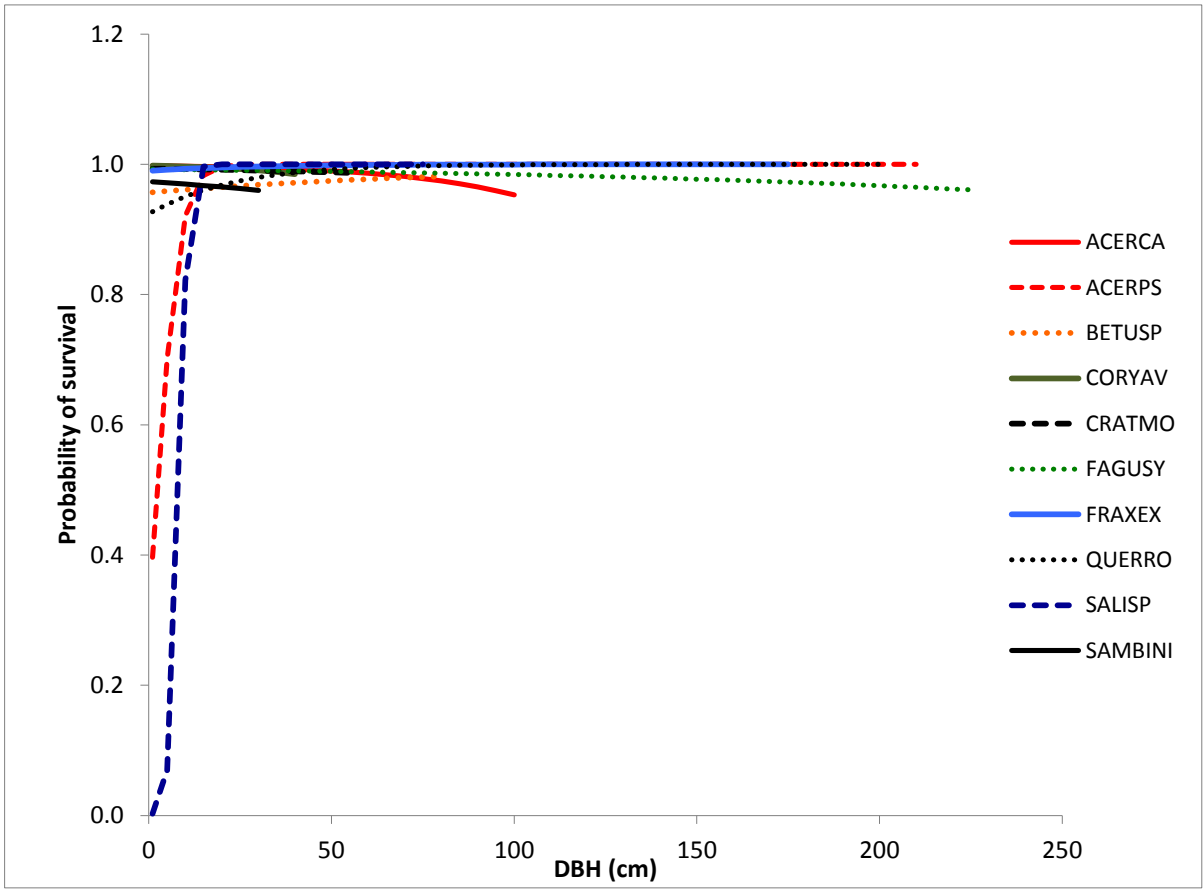

**Figure 4. Combined survival rates in all trees**

Probability of survival changes with DBH in trees at Wytham, derived from model 1 in table 2, which treats saplings and adults as equivalent but includes a DBH effect on all trees, the coefficients ($\beta_1$ and $\beta_2$) used to parameterise these functions are found in table 5. Data are plotted up to the maximum DBH recorded in Wytham.